\documentclass[final,1p,times]{elsarticle} 
\usepackage{graphicx} 
\usepackage{amssymb} 
\usepackage{amsthm} 
\usepackage{lineno} 
\def\be\begin{equation}
\def\ee\end{equation}
\journal{Nuclear Physics A} 
\begin{document} 

\begin{frontmatter} 


\title{The viscosity bound in string theory}

\author{Aninda Sinha and Robert C. Myers }

\address{Perimeter Institute for Theoretical Physics, Waterloo, Ontario N2L 2Y5, Canada}

\begin{abstract} 
The ratio of shear viscosity to entropy density $\eta/s$ of any
material in nature has been conjectured to have a lower bound of
$1/4\pi$, the famous KSS bound. We examine string theory models for
evidence in favour of and against this conjecture. We show that in a
broad class of models quantum corrections yield values of $\eta/s$
just above the KSS bound. However, incorporating matter fields in
the fundamental representation typically leads to violations of this
bound. We also
outline a program to extend AdS/CFT methods to RHIC phenomenology.\\
{\it Based on a talk given by AS at Quark Matter 2009, March 30 - April 4, Knoxville, Tennessee.}
\end{abstract} 

\end{frontmatter} 



\section{Introduction}

The AdS/CFT correspondence is one of the most profound developments
not only in string theory but also in the study of quantum field
theories in recent times. This holographic framework allows one to
use simple gravitational calculations to understand the strong
coupling behaviour of conformal gauge theories. Using this
correspondence, Kovtun, Son and Starinets showed that the ratio of
the shear viscosity to entropy density $\eta/s$ for a large class of
strongly coupled relativistic fluids satisfied \cite{kss}
\begin{equation}
\frac{\eta}{s}=\frac {1}{4\pi}\,. \label{kss}
\end{equation}
They went further to conjecture that this value  actually represents
a lower bound on this ratio for any fluid in nature, the KSS bound
\cite{kss}. While any known fluid respects this bound (by more than
an order of magnitude), RHIC data seems to indicate that the
strongly coupled quark-gluon plasma has $\eta/s\sim1/4\pi$
\cite{data}. This triggered a great deal of work to better
understand the bound and put it on a firmer footing. The KSS bound
is frequently referred to as the `quantum bound' \cite{rev}, since
this leading-order result corresponds to a regime where the 't Hooft
coupling $\lambda=g_{YM}^2 N_c$ and the number of colours $N_c$ are
both infinite. These limits are necessary in order to keep the
curvature corrections and string-loop corrections small in the
gravity calculations. Hence, it may seem rather curious that the
RHIC plasma seems to yield a value for $\eta/s$ quite close to
(\ref{kss}) but it may point to some type of universal behaviour. Of
course, the obvious question becomes can we compute $1/\lambda$ and
$1/N_c$ corrections -- which we will refer to as quantum corrections
-- to $\eta/s$?

This talk summarizes what is known from string theory about quantum
corrections to $\eta/s$. The punch-line is that in a wide class of
models where (\ref{kss}) is the leading result, the quantum
corrections are positive and so the bound is respected. However, a
common feature to all of these models is the absence of matter
fields in the fundamental representation, {\it i.e.,} `no quarks.'
It is also found that in many models with fundamental matter, the
leading corrections produce violations of the KSS bound. The
implications of these findings will be discussed in the conclusion.
This talk is based on \cite{mps08}-\cite{mps09}. For a more complete
list of references, please refer to \cite{rev}-\cite{mps09}.

\section{Quantum corrections to $\eta/s$}

For a 4d conformal gauge theory, the dual gravity action may be
written schematically as
\begin{equation}\label{act1}
I=\frac{1}{2\ell_P^3} \int d^5 x
\sqrt{-g}\,\left(\frac{12}{L^2}+R+L^2 \lambda_1 W^2+ L^4 \lambda_2
W^3+ L^6 \lambda_3 W^4+\cdots\right)\,.
\end{equation}
Here, $R$ and $W$ denote the Ricci scalar and Weyl tensor, respectively
for the 5d metric $g_{\mu\nu}$. Implicitly, we are working in a
perturbative expansion with respect to $\ell_P/L$, the ratio the 5d
Planck length to the AdS$_5$ curvature scale. The dimensionless
parameters controlling the higher curvature corrections are expected
to be $|\lambda_n|\sim (\ell_P/L)^{2n}\ll1$.  To describe a
situation with a nonvanishing chemical potentials, one would also
have to introduce a gauge field in the action. Further, there could
be scalar fields as well and in principle, one needs to examine how
these could affect the results. For the purpose of this talk, these
fields will be ignored, as may be rigorously justified in certain
approximations -- see \cite{mps08, bms08}. The precise tensor
contractions in $W^n$ as well as the values of $\lambda_n$'s are
string theory inputs. When the (zero temperature) gauge theory is
supersymmetric, $\lambda_2$ vanishes. Similarly, without fundamental
matter, the $\lambda_1$ term is absent. In this case, the leading
correction begins with the $\lambda_3$ term. From string theory, we
find that $\lambda_3=\zeta(3)/8\lambda^{3/2}$. It was shown
rigorously in \cite{mps08, bmps08} that starting with the full
ten-dimensional string theory, one indeed gets an action of the form
(\ref{act1}) with $\lambda_1=\lambda_2=0$. It was further confirmed
that there are no other fields in the full 10d solution which alter
the existing result \cite{bls} which is
\begin{equation}
\frac{\eta}{s}=\frac{1}{4\pi}\left(1+\frac{15\zeta(3)}{\lambda^{3/2}}
\right)\,.
\end{equation}
It was also argued in \cite{mps08} that the leading corrections in
the $1/N_c$ expansion appear at order $\lambda^{1/2}/N_c^2$ and are
also positive (again to emphasise, only in the absence of
fundamental matter).

To make a better comparison to real world QCD, one should also put
in fundamental matter and recalculate the leading correction. In
string theory, this requires more involved constructions involving
D7-branes and orientifold planes. These objects, in particular the
D7-branes, are known to produce a $W^2$ term in (\ref{act1}). While
a precise derivation of the effective 5d action becomes much more
complicated, we can invoke an indirect argument \cite{bms08} as
follows: Any four-dimensional conformal field theory can be
characterized by two central charges, $c$ and $a$, related to the
trace anomaly. The trace anomaly can be calculated using holographic
methods, yielding $L^3/\ell^3_P\simeq c/\pi^3$ and
$\lambda_1\simeq(c-a)/8c$ -- see \cite{bms08} for more details.
Thus if $c$ and $a$ are given by some knowledge of the gauge theory,
we can engineer the appropriate holographic action (\ref{act1})
which must arise from the full string theory. Then given the gravity
action, we can easily calculate \cite{oops}
\begin{equation}
\frac{\eta}{s}=\frac{1}{4\pi}(1-8\lambda_1 +
O(\lambda_3,\lambda_1^2))\,.
\end{equation}
(See \cite{mps09} for an efficient method of calculating the leading
corrections to $\eta/s$.) Recall that we are assuming a perturbative
expansion with $|\lambda_n| \ll 1$ in the gravity calculations. Thus
if and only if $\lambda_1<0$ will the leading correction be positive
and the KSS bound be respected. At this point, we can turn the
question around and ask if there are any superconformal gauge
theories for which we have both $c-a<0$ with $|c-a|/c \ll 1$.
Furthermore, for the next-to-leading order correction proportional
to $\lambda_3$ to be subdominant, we must also require $\lambda\gg
N_c^{2/3}$. One may have naively expected that there should be very
many such theories. However, in \cite{bms08}, we searched a whole
class of CFT's and found that not a single one satisfied these
criteria! All the examples we studied had $c-a>0$ indicating that
for these CFT's, $\eta/s$ would violate the KSS bound. For example,
all the models with gauge group $SU(N_c)$ and $|c-a|/c \ll 1$ as
$N_c\to \infty$, are shown in the following table.
\begin{table}[ht]
\centering
\begin{tabular}{|c|c|c|c|}
    \hline
      & $(n_{adj},n_{asym},n_{sym},n_f)$    &   $c-a$  & $\delta=(c-a)/c$     \\
    \hline
 (a)&(3,0,0,0) & 0 & 0   \\
    \hline
(b)&(2,1,0,2) & $\frac{3N_c+1}{48}$ & $\frac{1}{4N_c}+O(N_c^{-2})$   \\
    \hline
(c)&(1,2,0,4) & $\frac{3N_c+1}{24}$ & $\frac{1}{2N_c}+O(N_c^{-2})$   \\
    \hline
(d)&(1,1,1,0) & $\frac{1}{24}$ & $\frac{1}{6N_c^2}+O(N_c^{-4})$   \\
    \hline
(e)&(0,3,0,6) & $\frac{3N_c+1}{16}$ & $\frac{3}{4N_c}+O(N_c^{-2})$   \\
    \hline
(f)&(0,2,1,2) & $\frac{N_c+1}{16}$ & $\frac{1}{4N_c}+O(N_c^{-2})$   \\
    \hline
\end{tabular}
\caption{$\delta$ for $SU(N_c)$ models}
\end{table}
We should point out that this violation is $O(1/N_c)$ and hence very
small in the large $N_c$ limit (where the gravity calculations are
reliable). Note that in model (d), the correction from $\lambda_1$
is $1/N_c^2$ and hence subdominant compared to $\lambda_3$.
Therefore this example falls into the category of theories
preserving the bound, along with those with no fundamentals, {\it
i.e.,} $n_f=0$.

This brings us to the question as to what happens to this bound
violation if we added a chemical potential. In supergravity models,
one easy way to add a chemical potential is to turn on a R-charge.
This corresponds to adding a Maxwell term $-1/4 F_{ab} F^{ab}$ and
certain higher derivative $R F^2$ and $F^4$ terms to (\ref{act1}),
as was considered in \cite{mps09}. There, it was demonstrated that
turning on a R-charge chemical potential only worsens the violations
of the KSS bound.

\section{Conclusion}

While we have found counter-examples to the proposed KSS bound, it
is premature to conclude that there is no bound at all. In
\cite{blmsy}, it was argued that with Gauss-Bonnet gravity, where
$W^2$ is replaced by the Gauss-Bonnet term which makes the equations
of motion second order and removes any ghosts, one may consider the
coefficient $\lambda_1$ to be finite. However, demanding that the
dual gauge theory respect causality puts constraints on this
coefficient  and it was found that with these constraints in place
\cite{blmsy},
\begin{equation}
\frac{\eta}{s}\geq \frac{16}{25}\frac{1}{4\pi}
\end{equation}
for this class of models. It seems unlikely that this result
represents a fundamental bound. Hence it is interesting to speculate
as to whether a bound actually exists or whether $\eta/s$ can be
systematically reduced towards zero. In any event, holographic
constructions seem to provide an interesting new regime of fluid
dynamics.

What remains of course curious is the fact that the RHIC plasma has
a value for $\eta/s$ in the vicinity of $1/4\pi$. The current upper
bound sits at $\eta/s<0.2$ \cite{data}. Hence an interesting
question is if we can extend the AdS/CFT methods to do quantitative
phenomenology relevant for physics at RHIC or the LHC. Any progress
in this direction will be very worthwhile. One simple approach is
the following: Let us assume that there is fundamental matter and
the leading correction in (\ref{act1}) begins with $\lambda_1$. We
can show that the holographic energy density normalized by the free
field value \cite{bms08} is given by
\begin{equation}\label{eq1}
\frac{\varepsilon}{\varepsilon_0}=\frac{3}{4}\left(1+\frac{1}{4}\frac{c-a}{c}
\right) \equiv\frac{3}{4}\left(1+\frac{1}{4}\delta\right)\,,
\end{equation}
while
\begin{equation}\label{eq2}
\frac{\eta}{s}=\frac{1}{4\pi}\left(1-\frac{c-a}{c}\right)=
\frac{1}{4\pi}\left(1-\delta\right)\,.
\end{equation}
Current lattice calculations indicate that
$\varepsilon/\varepsilon_0$ is roughly between $0.85-0.90$. Plugging
this into (\ref{eq1}) yields $\delta\sim 0.53-0.80$, which using
(\ref{eq2}) then produces $\eta/s\sim 0.016-0.037$. This ratio is
clearly quite a bit lower than (\ref{kss}). However, quite
remarkably the consistency of the CFT's in general requires that
$|\delta|\le 0.5$ \cite{blmsy,const}. Thus, this exercise shows us
that in order to extend the AdS/CFT methods to RHIC phenomenology,
we will need to work harder. The next to leading correction which
begins with $\lambda_3$ in (\ref{act1}) for supersymmetric theories
will be equally important and to go beyond just qualitative results,
these need to be incorporated into the theory. When this is done, we
will have two parameters, $\lambda_1$ and $\lambda_3$, which we will
phenomenologically fix. This means we require more inputs from the
lattice and/or experiment to constrain the holographic model. Then
using these values, we will calculate $\eta/s$ to see if it yields a
sensible result or not. If it does, then using the same
phenomenological lagrangian, we will go on to calculate other
physical quantities, {\it e.g.,} the relaxation time, and see if
these can be seen to agree with experiments as well. This will be
one way of extending AdS/CFT methods to phenomenology. Of course,
for consistency as a string theory calculation, we will also need
$|\lambda_n|$'s to be small compared to unity. If we are lucky, then
the number of corrections to be put in will be manageable. So the
bottom line is that any numerology without incorporating higher
derivative corrections at least up to $\lambda_3$ in (\ref{act1})
should be taken with a grain of salt. It does seem like hard work to
get something sensible out of this endeavor.


\section*{Acknowledgments} 
Research at Perimeter Institute is supported by the Government of
Canada through Industry Canada and by the Province of Ontario
through the Ministry of Research \& Innovation. RCM also
acknowledges support from an NSERC Discovery grant and funding from
the Canadian Institute for Advanced Research.


\begin{thebibliography}{00} 
\bibitem{kss}
  P.~Kovtun, D.~T.~Son and A.~O.~Starinets,
  Phys.\ Rev.\ Lett.\  {\bf 94}, 111601 (2005)
  [arXiv:hep-th/0405231].

\bibitem{data}
D.~Teaney,
  Phys.\ Rev.\  C {\bf 68}, 034913 (2003)
  [arXiv:nucl-th/0301099];
M.~Luzum and P.~Romatschke,
  Phys.\ Rev.\  C {\bf 78}, 034915 (2008)
  [arXiv:0804.4015 [nucl-th]].

\bibitem{rev}
  T.~Schafer and D.~Teaney,
  arXiv:0904.3107 [hep-ph];
  D.~T.~Son and A.~O.~Starinets,
  Ann.\ Rev.\ Nucl.\ Part.\ Sci.\  {\bf 57}, 95 (2007)
  [arXiv:0704.0240 [hep-th]].

\bibitem{bls}
  A.~Buchel, J.~T.~Liu and A.~O.~Starinets,
  Nucl.\ Phys.\  B {\bf 707}, 56 (2005)
  [arXiv:hep-th/0406264];
  A.~Buchel,
  Nucl.\ Phys.\  B {\bf 803}, 166 (2008)
  [arXiv:0805.2683 [hep-th]].

\bibitem{mps08}
  R.~C.~Myers, M.~F.~Paulos and A.~Sinha,
  Phys.\ Rev.\  D {\bf 79}, 041901 (2009)
  [arXiv:0806.2156 [hep-th]].

\bibitem{bms08}
  A.~Buchel, R.~C.~Myers and A.~Sinha,
  JHEP {\bf 0903}, 084 (2009)
  [arXiv:0812.2521 [hep-th]].

\bibitem{bmps08}
  A.~Buchel, R.~C.~Myers, M.~F.~Paulos and A.~Sinha,
  Phys.\ Lett.\  B {\bf 669}, 364 (2008)
  [arXiv:0808.1837 [hep-th]].

\bibitem{mps09}
  R.~C.~Myers, M.~F.~Paulos and A.~Sinha,
  JHEP {\bf 0906}, 006 (2009)
  [arXiv:0903.2834 [hep-th]].

\bibitem{blmsy}
  M.~Brigante, H.~Liu, R.~C.~Myers, S.~Shenker and S.~Yaida,
  Phys.\ Rev.\ Lett.\  {\bf 100}, 191601 (2008)
  [arXiv:0802.3318 [hep-th]].

\bibitem{oops}
M.~Brigante, H.~Liu, R.~C.~Myers, S.~Shenker and S.~Yaida,
  Phys.\ Rev.\  D {\bf 77}, 126006 (2008)
  [arXiv:0712.0805 [hep-th]];
 Y.~Kats and P.~Petrov,
  JHEP {\bf 0901}, 044 (2009)
  [arXiv:0712.0743 [hep-th]].

\bibitem{const}
  D.~M.~Hofman and J.~Maldacena,
  JHEP {\bf 0805}, 012 (2008)
  [arXiv:0803.1467 [hep-th]];
D.~M.~Hofman,
  arXiv:0907.1625 [hep-th];
   A.~Buchel and R.~C.~Myers,
  arXiv:0906.2922 [hep-th].



\end{thebibliography}
\end{document}